\documentclass[twocolumn,superscriptaddress,aps,preprintnumbers,amsmath,amssymb,sort&compress,nofootinbib
]{revtex4-1}

\usepackage{amsfonts,amsmath,amssymb,ascmac,bm,tensor}
\usepackage{comment}
\usepackage{ifpdf}
\usepackage{slashed}
\usepackage{color}
\usepackage[mathscr]{eucal}
\usepackage[utopia]{mathdesign}
\usepackage[utf8]{inputenc}
 
\usepackage{cancel}

\ifpdf
  \usepackage{graphicx}     
  \usepackage[bookmarksopen,colorlinks=true,linkcolor=bblue,citecolor=bblue,urlcolor=ppink]{hyperref}
\else     
  \usepackage[dvipdfmx]{graphicx}     
  \usepackage[dvipdfmx,bookmarksopen,colorlinks=true,linkcolor=bblue,citecolor=ppink,urlcolor=darkred]{hyperref}
\fi

\definecolor{red}{rgb}{1,0,0}
\definecolor{darkred}{rgb}{0.6,0,0}
\definecolor{darkgreen}{rgb}{0.992447,0.623778,0.034597}
\definecolor{ppink}{rgb}{1,0.4,0.4}
\definecolor{bblue}{rgb}{0.284602,0.317763,0.963947}

\newcommand{\1}{\mbox{1}\hspace{-0.25em}\mbox{l}}
\newcommand{\vev}[1]{ \left< {#1} \right> }

\newcommand{\prn}[1]{\left( {#1} \right)}

\newcommand{\dd}{\mathrm{d}}

\newcommand{\abs}[1]{\left\vert {#1} \right\vert}

\newcommand{\ee}{\mathrm{e}}

\makeatletter
\newcommand\footnoteref[1]{\protected@xdef\@thefnmark{\ref{#1}}\@footnotemark}
\makeatother

\begin{document}


\title{
Double inflation as a single origin of primordial black holes for all dark matter and LIGO observations
}
\author{Keisuke Inomata}
\affiliation{ICRR, University of Tokyo, Kashiwa, 277-8582, Japan}
\affiliation{Kavli IPMU (WPI), UTIAS, University of Tokyo, Kashiwa, 277-8583, Japan}
\author{Masahiro Kawasaki}
\affiliation{ICRR, University of Tokyo, Kashiwa, 277-8582, Japan}
\affiliation{Kavli IPMU (WPI), UTIAS, University of Tokyo, Kashiwa, 277-8583, Japan}
\author{Kyohei Mukaida}
\affiliation{DESY, Notkestra{\ss}e 85, D-22607 Hamburg, Germany}
\author{Tsutomu T.~Yanagida}
\affiliation{Kavli IPMU (WPI), UTIAS, University of Tokyo, Kashiwa, 277-8583, Japan}
\affiliation{Hamamatsu Professor}

\begin{abstract}
\noindent
Primordial Black Hole (PBH) is one of the leading non-particle candidates 
for dark matter (DM).
Although several observations severely constrain the amount of PBHs,
it is recently pointed out that there is an uncertainty on the microlensing constraints
below $\sim 10^{-10} M_\odot$ which was ignored originally
but may weaken the constraints significantly.
In this paper, facing this uncertainty, we investigate the possibility that 
PBHs can make up all DM in a broad mass spectrum.
Moreover, we propose a concrete inflation model which can simultaneously produce
PBHs for all DM in a broad mass spectrum around $\mathcal O(10^{-13}) M_\odot$ 
and PBHs for LIGO events in a sharp mass spectrum at $\mathcal O(10) M_\odot$.

\end{abstract}

\date{\today}
\maketitle
\preprint{IPMU 17-0158}
\preprint{DESY 17-190}

\section{Introduction}\label{sec: intro}

The identity of dark matter (DM) is one of the remaining mysteries of physics.
Although the existence of DM has been confirmed by the astrophysical and cosmological observations,
the property is almost unclear, except that it interacts at least gravitationally.
Common ways to explain DM involve new particles predicted in the beyond standard model;
such as weakly interacting massive particle and axion.
Another approach incorporates astronomical objects, such as black holes (BHs).
In the latter case, we need not introduce new elementary particles for DM.
Hence, it is important to rethink the possibility that such objects can make up all DM even from the viewpoint of high-energy physics.

The observations of the cosmic microwave background (CMB) set an obvious restriction on the astronomical objects as DM because they should behave as DM in the recombination era, which is much before ordinary stellar objects are formed.
Since primordial black holes (PBHs)~\cite{Hawking:1971ei,Carr:1974nx,Carr:1975qj} can be generated even at the radiation-dominated era, they are primary candidates for the astronomical objects as DM.  
Although the mass of PBHs can vary in many orders of magnitude in principle, several observations have ruled out most of the region for PBH DM.
Recently, however, the authors of the microlensing observation with Subaru Hyper Supreme-Cam (HSC)~\cite{Niikura:2017zjd} have mentioned that
the wave effect may weaken the constraints below $\sim 10^{-10}M_\odot$ significantly, which was ignored originally~\cite{Niikura:2017zjd,Takada:2017sympo,Takada:2017pbhfocus}.
Therefore the PBHs may still have the potential to explain all DM at least
around $10^{-14} M_\odot$--$10^{-10} M_\odot$.\footnote{
	PBHs in this mass range are also interesting
	since they can account for $r$-process nucleosynthesis~\cite{Fuller:2017uyd}.
}

Meanwhile, LIGO-Virgo collaborations have detected several gravitational waves (GWs)
 and discovered Black Holes (BHs) and Neutron Star(s) which produce the GWs through their mergers~\cite{Abbott:2016blz,Abbott:2016nmj,Abbott:2017vtc,Abbott:2017oio,TheLIGOScientific:2017qsa,Abbott:2017gyy}.
Table~\ref{tab:mass_summary} shows the masses of the observed BHs.
From this table, we can see that the five BHs out of twelve BHs detected by LIGO-Virgo collaborations have the masses around $30 M_\odot$.
While the stellar origin BHs produced in the usual metallicity environment ($Z\sim Z_\odot$) may not be as heavy as $30 M_\odot$~\cite{TheLIGOScientific:2016htt,Belczynski:2009xy,Spera:2015vkd},
PBHs can have such masses because their formation mechanism is completely different.
Therefore PBH is thought to be one of the candidates for such BHs~\cite{Bird:2016dcv,Clesse:2016vqa,Sasaki:2016jop,Eroshenko:2016hmn,Carr:2016drx}.

\begin{table}[t]
\begin{tabular}{c||c|c}
&\,Primary mass\, &\, Secondary mass\\
\hline
GW150914\, &  $36.2\,^{+5.2}_{-3.8}$\,$M_\odot$ & $29.1\,^{+3.7}_{-4.4}$\,$M_\odot$\\
[.3em]
\hline
LVT151012\, & $23\,^{+18}_{-6}$\,$M_\odot$  & $13\,^{+4}_{-5}$\,$M_\odot$ \\
[.3em]
\hline
GW151226\, &  $14.2\,^{+8.3}_{-3.7}$\,$M_\odot$  & $7.5\,^{+2.3}_{-2.3}$\,$M_\odot$ \\
[.3em]
\hline
GW170104\, & $31.2\,^{+8.4}_{-6.0}$\,$M_\odot$   & $19.4\,^{+5.3}_{-5.9}$\,$M_\odot$  \\
[.3em]
\hline
GW170608\, & $12\,^{+7}_{-2}$\,$M_\odot$   & $7\,^{+2}_{-2}$\,$M_\odot$  \\
[.3em]
\hline
GW170814\, & $30.5\,^{+5.7}_{-3.0}$\,$M_\odot$   & $25.3\,^{+2.8}_{-4.2}$\,$M_\odot$  \\
\end{tabular}
\caption{\small The masses of the BHs detected by LIGO-Virgo collaborations~\cite{Abbott:2016blz,Abbott:2016nmj,TheLIGOScientific:2016pea,Abbott:2017vtc,Abbott:2017oio,Abbott:2017gyy}.
}
\label{tab:mass_summary}
\end{table}

The production of PBHs requires an over-dense region 
that can overcome the radiation pressure.
Such an over-dense region may originate from the primordial perturbations produced during inflationary era.
Roughly speaking,
if an inflaton experiences a plateau potential during inflation, 
the produced perturbations become large.
Later, the large perturbations collapse to PBHs at their horizon reentry~\cite{GarciaBellido:1996qt,Kawasaki:1997ju,Yokoyama:1998pt}.
Therefore, the mass function of PBHs depends on the power spectrum of the curvature perturbations, that is, the properties of inflation models.

In general, any realistic inflation model results in the extended mass function,
not a monochromatic mass function.
As discussed in~\cite{Carr:2016drx,Green:2016xgy,Inomata:2017okj,Kuhnel:2017pwq,Carr:2017jsz,Bellomo:2017zsr}, 
extended mass functions are constrained more severely than monochromatic ones.
To narrow down inflation models for PBH DM,
it is quite important to know observationally how broad the mass function can be.
The slow-roll parameters determine
the spectral tilt of the curvature perturbations which exit the horizon during \textit{slow-roll} inflation. 
As a result, the perturbations produced during the slow-roll inflation predict the \textit{broad} mass spectrums~\cite{Clesse:2015wea,Kawasaki:2016ijp,Kawasaki:2016pql}.
On the other hand, some inflation models violate the slow-roll conditions in the inflationary era and realize the \textit{sharp} mass spectrums~\cite{Inomata:2016rbd,Inomata:2017okj,Kannike:2017bxn,Ballesteros:2017fsr}.

In this paper, taking account of the uncertainty of the microlensing constraints,
we revisit the idea of PBHs as all DM and PBHs for LIGO events.
We show that PBHs can constitute all DM even in a broad mass spectrum at $10^{-14} M_\odot$--$10^{-10} M_\odot$.
Interestingly, this uncertainty opens up a possibility that
the double inflation model can account for PBHs for all DM and for LIGO at once
because it can have two peaks in the curvature perturbation, sharp and broad ones.
We explicitly show this is actually possible 
by identifying a sharp peak as PBHs for LIGO and a broad one as those for all DM,
and discuss its implication on the induced GWs via the second order effects.

\section{PBH formation}
\label{sec:pbh_formation}

In this section, we briefly review the basic formulae for the formation of PBHs (see also~\cite{Josan:2009qn,Carr:2016drx}).
Throughout this paper,
we consider the PBHs produced in the radiation dominated era.\footnote{
PBHs produced in the matter dominated era are discussed in~\cite{Khlopov:1980mg,Khlopov:1982sov,Harada:2016mhb}.}
If the perturbations are large enough, the gravity of the over-dense regions can overcome the pressure force of radiation and then collapse to form BHs soon after the horizon entry.
According to Carr~\cite{Carr:1975qj}, the threshold of the density perturbations for the PBH formation is estimated as $\delta_c \simeq 1/3$ by the simple analysis. 
Recent numerical and analytical studies suggest the threshold value of $\delta_c \simeq 0.4$~\cite{Musco:2012au,Harada:2013epa}. 
We adopt $\delta_c \simeq 0.4$ as a fiducial value for the threshold.
The mass of a PBH is related to the horizon mass at the horizon entry of the perturbation.
The relation between the scale of the perturbation and 
the PBH mass is given by 
\begin{align}
	M
	&= \left. \gamma \rho \frac{4 \pi H^{-3}}{3} \right|_{k = aH}
	\simeq\frac{\gamma M_\text{eq}}{\sqrt{2}} 
	\prn{ \frac{g_{\ast,\text{eq}}}{g_\ast} }^\frac{1}{6}
	\prn{ \frac{k_\text{eq}}{k} }^2  \nonumber \\[.5em]
	&\simeq M_\odot\left(\frac{\gamma}{0.2}\right)\left(\frac{g_*}{10.75}\right)^{-\frac{1}{6}}\left(\frac{k}{1.9\times10^6\,\mathrm{Mpc}^{-1}}\right)^{-2} 
	\label{eq:pbhmass in k} \\[.5em]
		&\simeq 10^{-13}\,M_\odot\,
    \left(
    \frac{\gamma}{0.2}
    \right)
    \left(
    \frac{g_\ast}{106.75}
    \right)^{- \frac{1}{6}}
    \left(
    \frac{k}{4.9\times 10^{12} \,\textrm{Mpc}^{-1}}
    \right)^{-2} \\[.5em]
	&\simeq M_\odot
    \left(
    \frac{\gamma}{0.2}
    \right)
    \left(
    \frac{g_\ast}{10.75}
    \right)^{- \frac{1}{6}}
    \left(
    \frac{f}{2.9 \times 10^{-9} \,\textrm{Hz
   }}
    \right)^{-2} \label{eq:pbhmass_freq}\\[.5em]
    &\simeq 10^{-13} \,M_\odot \,
    \left(
    \frac{\gamma}{0.2}
    \right)
    \left(
    \frac{g_\ast}{106.75}
    \right)^{- \frac{1}{6}}
    \left(
    \frac{f}{7.8 \times 10^{-3} \,\textrm{Hz
   }}
    \right)^{-2},
    	\label{eq:pbhmass}
\end{align}
where we have also estimated the corresponding frequency, $f\equiv k/2\pi$ for later convenience.
$\gamma$ is the fraction of the PBH mass in the horizon mass at the formation and depends on the detail of the gravitational collapse.
$\gamma$ is estimated as $\gamma\sim(1/\sqrt{3})^3$ by the simple analytical calculation~\cite{Carr:1975qj}
 and we adopt this value as a fiducial value in the following.
$g_*$ is the effective degrees of freedom.
The subscript ``eq'' means the value at the matter-radiation equality time, 
and in particular, $M_\text{eq}$ is the horizon mass at the equality time.

The production rate of PBH can be interpreted as the appearance rate of the perturbations larger than the threshold.
Assuming that the perturbations follow the Gaussian distribution\footnote{
We focus on the case where the curvature perturbations follow the Gaussian distribution throughout this paper.
The effects of non-Gaussianity for PBH formation are discussed in~\cite{Byrnes:2012yx,Young:2013oia,Nakama:2016kfq}.
},
the production rate of PBH can be expressed as
\begin{align}
	\beta (M) =
	\int_{\delta_c}
	\frac{\dd \delta}{\sqrt{2 \pi \sigma^2 (M)}} \, e^{- \frac{\delta^2}{2 \sigma^2 (M)}}
	\simeq 
	\frac{1}{\sqrt{2 \pi}} \frac{1}{\delta_c / \sigma (M)} \, e^{- \frac{\delta_c^2}{2 \sigma^2 (M)}}.
	\label{eq:beta}
\end{align}
$\sigma(M)^2$ is the coarse-grained density contrast with the smoothing scale $k$, which is defined as~\cite{Young:2014ana}
\begin{align}
	\sigma^2 (M (k))
	= \int \dd \ln q W^2 (q k^{-1}) \frac{16}{81} \prn{q k^{-1}}^4 \mathcal P_\zeta (q),
	\label{eq:sigma}
\end{align}
where $\mathcal P_\zeta(k)$ is the power spectrum of the curvature perturbations.
$W(x)$ is the window function and we take the Gaussian filter 
defined as $W(x) = \ee^{-x^2/2}$.

The fraction of PBHs for DM is often used to compare the theoretical prediction of the PBH abundance with the observational results. Using the production rate, $\beta(M)$, we derive the following formula for the PBH DM fraction:
\begin{align}
	&f(M)
	\simeq  \left. \frac{\rho_\text{PBH}(M)}{\rho_m} \right|_\text{eq} \frac{\Omega_{m}}{\Omega_\text{DM}}
	= \prn{\frac{T_M}{T_\text{eq}} \frac{\Omega_{m}}{\Omega_\text{DM}}} \gamma \beta (M) \nonumber \\
	& \simeq
	\!\left(\! \frac{\beta (M)}{1.84 \times 10^{-8}} \!\right)
	\!\left(\! \frac{\gamma}{0.2} \!\right)^\frac{3}{2}\!
	\!\left(\! \frac{10.75}{g_{\ast} (T_M)} \!\right)^\frac{1}{4}\!
	\!\left(\! \frac{0.12}{\Omega_\text{DM} h^2} \!\right)
	\!\left(\! \frac{M}{M_\odot} \!\right)^{-\frac{1}{2}} \hspace{-7pt} \label{eq:frac1}\\
	&\simeq
	\prn{ \frac{\beta (M)}{1.04 \times 10^{-14}} }
	\prn{\frac{\gamma}{0.2}}^\frac{3}{2}
	\prn{ \frac{106.75}{g_{\ast} (T_M)} }^\frac{1}{4}
	\prn{\frac{0.12}{\Omega_\text{DM} h^2}}
	\prn{ \frac{M}{10^{-13}\,M_\odot} }^{-\frac{1}{2}},
	\label{eq:frac2}
\end{align}
where $f(M) \equiv \frac{1}{\Omega_\text{DM}} \frac{\dd\, \Omega_\text{PBH}}{\dd\, \text{ln}\,M}$ and $\rho_\text{PBH}(M)\equiv \frac{\dd\, \rho_\text{PBH}}{ \dd\, \text{ln} \, M}$
 are the differential mass function of the PBH DM fraction and the PBH energy density, respectively.
The subscripts ``$m$'' and ``DM'' mean the matter (baryon + DM) and DM (DM only), with $\Omega_\text{DM} h^2 \simeq 0.12$~\cite{Ade:2015xua}.
$T_M$ represents the temperature at which the PBHs of the mass $M$ are produced.
From Eq.~(\ref{eq:frac1}) and (\ref{eq:frac2}), we can estimate  typical curvature perturbations to generate a sizable amount of PBHs as $\mathcal P_\zeta \sim \mathcal O(0.01)$, 
which is much larger than that on the CMB scale ($\mathcal P_\zeta \sim \mathcal O(10^{-9})$ at $k\lesssim \mathcal O(1)$Mpc$^{-1}$~\cite{Ade:2015xua,Nicholson:2009new,Nicholson:2010new,bird2011minimally}).
The total fraction of PBHs in DM is given by
\begin{align}
	\frac{\Omega_{\text{PBH,tot}}}{\Omega_\text{DM}} = \int \dd \ln M\, f(M).
\end{align}

\section{Constraints on PBH abundance}
\label{sec:constraints}

In this section, we summarize the constraints on the abundance of PBHs  
whose masse are $\mathcal O(10^{-14}$--$10^{-10}) M_\odot$
and $\mathcal O(10) M_\odot$ and discuss their uncertainties.

\paragraph*{\bf Constraints on $\mathcal O(10^{-14}$--$10^{-10}) M_\odot$ PBHs.}

This mass region is mostly constrained by the microlensing observation with Subaru HSC~\cite{Niikura:2017zjd} 
and the observation of the remaining white dwarfs~\cite{Graham:2015apa}.

The gravitational microlensing occurs when the lens-objects pass through our line of sight to background stars 
and is observed as the temporary amplification of the light of the background stars.
Here, we mention the current status of microlensing constraints.
MACHO/EROS/OGLE experiments set constraints on the abundance of the massive compact halo objects (MACHOs), including PBHs, with $[10^{-7} , 10]M_\odot$
 observing the Large Magellanic Cloud (LMC) and the Small Magellanic Cloud (SMC), $\sim 50$ kpc and $60$ kpc~\cite{Allsman:2000kg,Tisserand:2006zx,wyrzykowski2011ogle}.
Meanwhile, Griest \emph{et al.}\ have constrained the abundance of PBHs with $\mathcal O(10^{-8})M_\odot$ with the use of the data from Kepler satellite~\cite{Griest:2013esa}.
Recently, Niikura \emph{et al.}\ have put severe constraints on the PBH abundance with $[10^{-13} , 10^{-6}]M_\odot$ observing the stars in the Andromeda galaxy (M31: $\sim 1$Mpc) with the Subaru HSC~\cite{Niikura:2017zjd}.
The reason why the constraint covers the light mass range is due to its high-cadence (HSC: $2$ min sampling, Kepler: $30$ min sampling, EROS/MACHO/OGLE: $\mathcal O(10)$ min--$\mathcal O(1)$ day sampling).

The result of HSC severely constrains the PBH DM scenario around the sublunar mass.
However, there is a phenomenon that they have mentioned, but not taken into account in their current analysis~\cite{Niikura:2017zjd,Takada:2017sympo,Takada:2017pbhfocus}.
That is so-called ``wave effect''.
The theory of the amplification due to the gravitational microlensing is based on the geometrical optics approximation.
However, when the wavelength of light is larger than the Schwarzschild radius of the lensing object, the diffraction due to the wave properties of light should be taken into account
 and the geometrical optics approximation becomes invalid~\cite{Ohanian:1974ys,Bliokh:1975wave,Bonts:1981wave}.
The diffraction of light lowers the maximum magnification of the microlensing signal~\cite{Nakamura:1997sw,Takahashi:2003ix}.
Since the observational wavelength of the HSC is $\sim 600$ nm ($r$-band),
 the constraints on PBHs with $\lesssim 2 \times 10^{-10}M_\odot$ ($=4\times 10^{23}$g) are expected to be weakened.\footnote{
In addition to the wave effect, they have mentioned another uncertainty related to the so-called ``finite source size effect''~\cite{Niikura:2017zjd}.
This uncertainty could also possibly weaken the constraint.
 }
Therefore, in Sec.~\ref{sec:inflation_model}, 
we 
vary the critical mass of PBH below which there are no constraints
owing to  the diffraction,
and see how the constraints on the extended mass function change.
To be concrete, we consider the following two cases:
no constraints for $\lesssim 2 \times 10^{-10}M_\odot$, and
no constraint for $\lesssim 2 \times 10^{-11}M_\odot$.

In addition to the constraint from the microlensing observation,
there is another constraint on PBHs around the sublunar mass 
from the observation of white dwarfs~\cite{Graham:2015apa}.
When the PBHs go through white dwarfs, the white dwarfs get heated by the dynamical friction of PBHs.
If the heat is large enough, the fusion reaction occurs in the components of the white dwarf (such as carbon), the heat from the fusion triggers the more fusion reactions, and finally the white dwarf explodes as a supernova.
Hence the observations of remaining white dwarfs can constrain the abundance of PBHs with $[10^{-14} , 10^{-13}]M_\odot$. 
However, note that this constraint includes the uncertainties about the physics of the dynamical heating and induced white dwarf explosion.
In addition, when they constrain the PBH abundance, they do not take into account the Poisson statistics, which is taken into account in Refs.~\cite{Allsman:2000kg,Tisserand:2006zx,wyrzykowski2011ogle,Niikura:2017zjd}.
Therefore, in Sec.~\ref{sec:inflation_model}, we consider both cases where we adopt the constraint from the white dwarfs, and where we neglect the constraint because of their uncertainties.

On the lighter mass range (<$\mathcal O(10^{-14}) M_\odot$), 
there are constraints from the observations of the femtolensing events~\cite{Barnacka:2012bm}
and the extragalactic gamma-ray produced by the Hawking evaporation of the light PBHs~\cite{Carr:2009jm}.

\paragraph*{\bf Constraints on $\mathcal O(10) M_\odot$ PBHs.}

The current microlensing observations do not severely constrain the abundance of $\mathcal O(10) M_\odot$ PBHs.
However, in this mass range, there are many other probes:
the variation of CMB spectrum~\cite{Poulin:2017bwe,Ali-Haimoud:2016mbv},
the radio and X-ray from accretion~\cite{Inoue:2017csr,Gaggero:2016dpq},
the dynamical heating of dwarf galaxies~\cite{Koushiappas:2017chw} and ultra-faint dwarf galaxies~\cite{Brandt:2016aco}, 
and the distribution of wide binaries~\cite{monroy2014end}.
If we take all these constraints, the abundance of $\mathcal O(10) M_\odot$ PBHs is constrained $\Omega_\text{PBH}/\Omega_\text{DM} \lesssim \mathcal O(10^{-2})$ ~\cite{Carr:2017jsz}.
Although these constraints may have astrophysical uncertainties individually, 
it seems that eluding all of them would require some special mechanism.
In this sense, it is difficult to explain all the DM by PBHs with $\mathcal O(10)M_\odot$.
Moreover, Sasaki \emph{et al.}\ have shown that, in order for PBHs to reproduce the merger rate expected by LIGO-Virgo collaborations ($12$-$213$\,Gpc$^{-3}$yr$^{-1}$~\cite{Abbott:2017vtc}),
the abundance of PBHs should be $\Omega_\text{PBH}/\Omega_\text{DM} \sim \mathcal O(10^{-3})$--$\mathcal O(10^{-2})$~\cite{Sasaki:2016jop}.
Similar analyses have been performed recently by \cite{Ali-Haimoud:2017rtz} and reached the same conclusion.

Once we specify the production mechanism of PBHs,
there are several indirect constraints on the PBH abundance.
In the following, we assume inflation as an origin of PBHs.
As we explained in Sec.~\ref{sec:pbh_formation}, the sizable amount of PBHs is realized by the large primordial perturbations. 
Although the typical probability of forming PBHs is extremely small,
there are huge numbers of over densities at the horizon reentry of the large perturbation.
These regions fail to collapse into PBHs, but we can use them to probe PBHs from large primordial density perturbations.
One method is to utilize the GWs induced by such large density perturbations via the second order effect.
Those GWs are constrained currently by pulsar timing array (PTA) experiments~\cite{Saito:2008jc,Saito:2009jt,Bugaev:2009zh,Bugaev:2010bb}.
In particular, in the context of PBHs for LIGO events, the PTA constraints have been discussed in Ref.~\cite{Inomata:2016rbd}, which is followed by Refs.~\cite{Orlofsky:2016vbd,Nakama:2016gzw}.
The frequency constrained by the PTA experiments, $f\sim$nHz, corresponds to the mass range of $\mathcal O(1) M_\odot$ PBHs (see Eq.~(\ref{eq:pbhmass_freq})).
Another way is to look at how these perturbations are dissipated into the background thermal plasma.
Sensitive probes depend on the era when the perturbed region reenters the horizon:
CMB spectral distortions ($\mu$- and $y$-distortions) by COBE/FIRAS at $k\sim 1$--$10^{4}$\,Mpc$^{-1}$~\cite{Fixsen:1996nj,Chluba:2012we,Kohri:2014lza},
and the change of big-bang nucleosinthesis (BBN) at $k\sim 10^{4}$--$10^{5}$\,Mpc$^{-1}$~\cite{Jeong:2014gna,Nakama:2014vla,Inomata:2016uip}.
The scale constrained by the COBE/FIRAS and BBN, $k<\mathcal O(10^5)$Mpc$^{-1}$, corresponds to $>\mathcal O(100) M_\odot$ (see Eq.~(\ref{eq:pbhmass in k})).
These indirect constraints are so severe that the curvature perturbations must damp quickly 
both above and below the scale corresponding to $\mathcal O(10) M_\odot$.
In other words, the peak of the curvature perturbations at the $\mathcal O(10) M_\odot$ scale should be sharp.\footnote{
If there is a sizable non-Gaussianity, the constraints from the PTA experiments and $\mu$-distortion observation can be weakened~\cite{Nakama:2016kfq,Nakama:2016gzw,Nakama:2017xvq,Ando:2017}. 
}

\section{Constraints on extended mass spectrums}
\label{sec:pbh_as_dm}

As discussed in~\cite{Carr:2016drx,Green:2016xgy,Inomata:2017okj,Kuhnel:2017pwq,Carr:2017jsz,Bellomo:2017zsr}, 
the constraints on the extended PBH mass spectrums can be more severe than those on the monochromatic ones.
Most of the observational constraints are based on the assumption that the PBH mass spectrum is monochromatic.
However, in reality, the mass spectrums are expected to have a finite width
and therefore the careful treatments are needed when we compare theoretical predictions with the observational constraints.

Here, let us review the analysis of the constraints on the extended mass function, which is discussed in Refs.~\cite{Inomata:2017okj,Carr:2017jsz}.
We define the astrophysical observable related to PBHs as $A[f(M)]$.
In general, $A[f(M)]$ can be expanded in terms of $f(M)$ as
\begin{align}
&A[f(M)] = A_0 + \int \dd\, \text{ln}\, M f(M) K_1(M) \nonumber \\
&+ \int \dd\, \text{ln}\, M_1 \dd\, \text{ln}\, M_2 f(M_1)f(M_2) K_2(M_1,M_2)+ \cdots ,
\label{eq:A_expand}
\end{align}
where $A_0$ is the background contribution and $K_j$ depends on the properties of the observable.
Most observables are determined by the terms up to the $K_1(M)$ term~\cite{Carr:2017jsz}.
Hence we neglect the higher order contributions to the observables in the following discussion.
In the case of the monochromatic mass function, it is given by
\begin{align}
	f_\text{mono}(M) \equiv f_\text{norm}(M_c) \, \delta( \text{ln}\,M -  \text{ln}\,M_c),
\end{align}
where $f_\text{norm}$ determines the normalization of the monochromatic function.
The constraints on $f_\text{norm}(M_c)$ is related to the upper bound of the observable, $A_\text{obs}$, as follows:
\begin{align}
	&A[f_\text{mono}(M)] < A_\text{obs}, \nonumber \\
	\Rightarrow \quad
	&A_0 + f_\text{norm}(M_c) K_1(M_c) < A_\text{obs}, \nonumber \\
	\Rightarrow \quad
	& f_\text{norm}(M_c) < \frac{A_\text{obs} - A_0}{K_1(M_c)} \equiv f_\text{obs}(M_c),
	\label{eq:f_max_def}
\end{align}
where we have defined $f_\text{obs}(M_c)$ as the observational upper bound of $f_\text{norm}(M_c)$.
From Eqs.~(\ref{eq:A_expand})--(\ref{eq:f_max_def}), we get the following inequality:
\begin{align}
	\int \dd\, \text{ln}\, M \frac{f(M)}{f_\text{obs}(M)} \leq 1.
	\label{eq:extend_cond}
\end{align}
This is the condition that the extended mass function should satisfy.
We use this condition in Sec.~\ref{sec:inflation_model} to check that our predicted mass function is consistent with the observational constraints.

\begin{figure*}
  \begin{minipage}[b]{0.49\linewidth}
    \centering
    \includegraphics[keepaspectratio, scale=0.5]{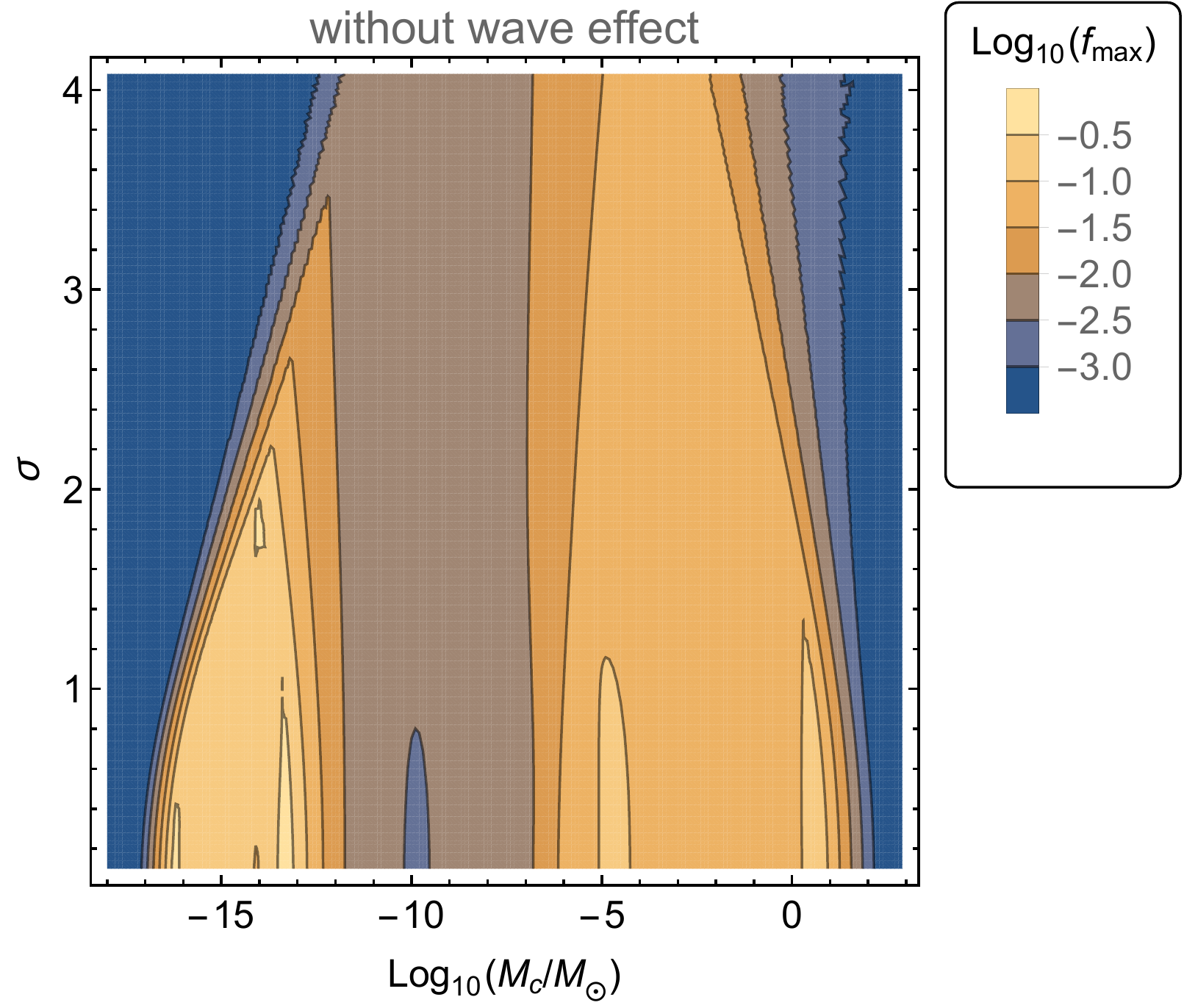}
  \end{minipage}
  \begin{minipage}[b]{0.49\linewidth}
    \centering
    \includegraphics[keepaspectratio, scale=0.5]{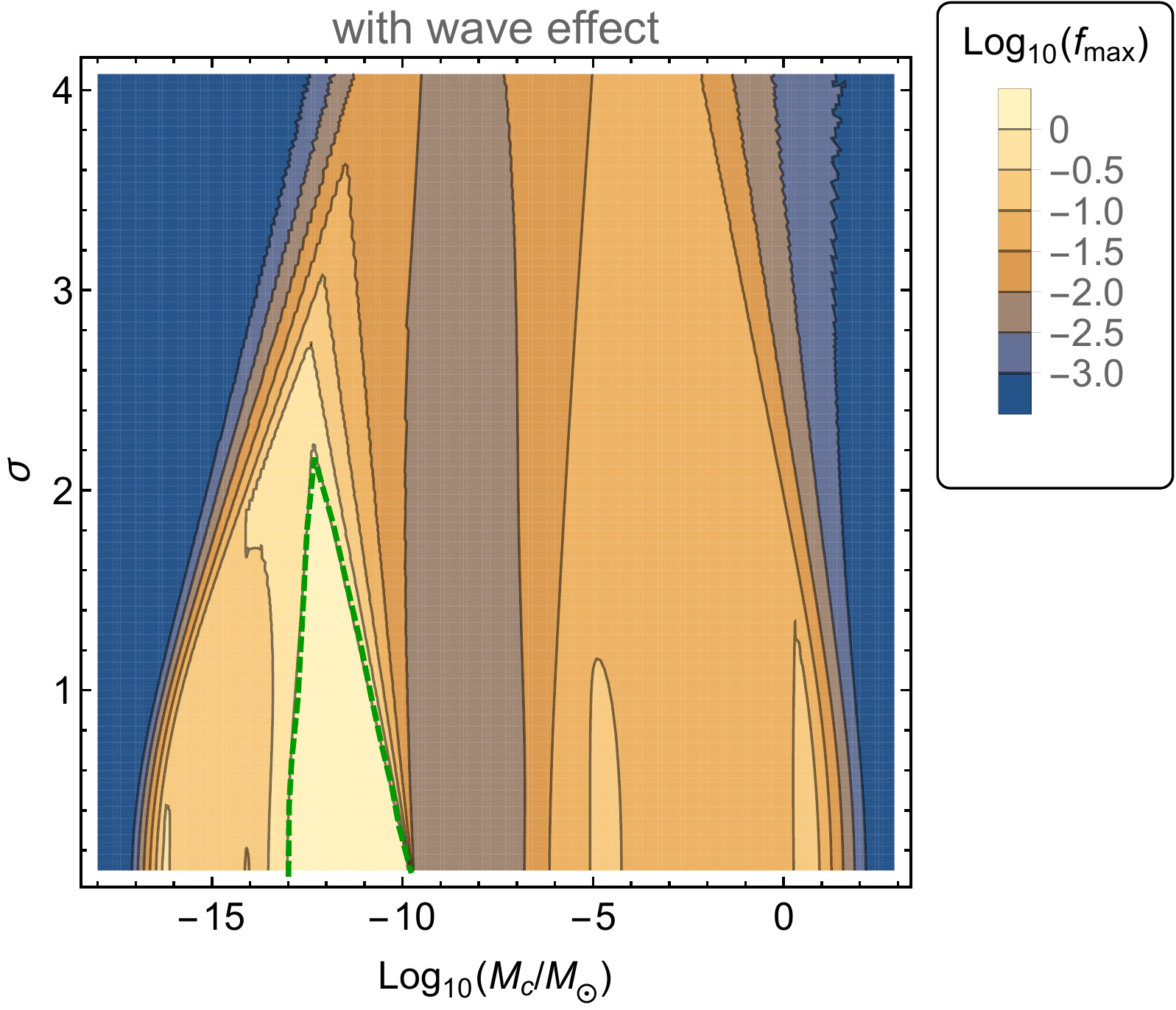}
  \end{minipage}
  \caption{\small	
	The constraints on the parameters of the extended mass function defined by Eq.~(\ref{eq:para_extend_f}).
	We use the monochromatic constraints shown in the Fig.~\ref{fig:pbh_spectrum} (including the constraint from the white dwarfs) to derive this result.
	In each point of ($M_c, \sigma$), we calculate the maximum value of $f_\text{max}$ under the condition of Eq.~(\ref{eq:extend_cond}).
	The left figure shows the result in the case where we adopt all the constraints
	 and the right figure shows the result in the case where we take into account the uncertainty related to the wave effect
         and neglect the HSC constraints on $<2 \times 10^{-10}M_\odot$ PBHs.
         A Green dashed line in the right figure shows the line of $f_\text{max} =1$, inside which PBHs can be all the DM.
	}	
	\label{fig:dm_scat}
\end{figure*}

Fig.~\ref{fig:dm_scat} shows the constraints on the parameters of the extended mass function defined as
\begin{align}
	f(M) = \frac{f_\text{max}}{\sqrt{2\pi} \sigma} \text{exp}\left[ -\frac{(\text{ln}(M/M_c))^2}{2\sigma^2} \right].
	\label{eq:para_extend_f}
\end{align}
In each point of ($M_c, \sigma$), we calculate the maximum value of $f_\text{max}$ under the condition of Eq.~(\ref{eq:extend_cond}).
From this figure, one can see that if the HSC constraints are weakened by the wave effect, 
there appears an open window for the PBHs as DM with $\mathcal O(10^{-13})M_\odot$.
Moreover, in that case, the PBHs as DM can be realized with the broad spectrum, 
e.g. the spectrum defined in Eq.~(\ref{eq:para_extend_f}) with $\sigma \simeq 2$ and $M_c \simeq 4\times 10^{-13}M_\odot$.

\section{Concrete inflation model}
\label{sec:inflation_model}

Now we are in a position to discuss the double inflation model~\cite{Kawasaki:1997ju} as a concrete example 
and show that this model can simultaneously produce the $\mathcal O(10^{-13})M_\odot$ PBHs explaining all the DM and the $\mathcal O(10)M_\odot$ PBHs explaining LIGO events.

The double inflation model has two stages of inflation, pre-inflation and new-inflation.
In this paper, we consider the following potentials:
(throughout this paper, we set $M_\text{Pl}=1$)
\begin{align}
	\label{eq:inf_potential_tot}
	V(\varphi,\chi) 
	&= V_\text{pre} (\chi) + V_\text{new} (\varphi) + V_\text{stb} (\varphi,\chi), \\
	\label{eq:inf_potential_new1}
	V_\text{new} (\varphi) 
	&= -2\sqrt{2} c v^2 \varphi - \frac{\kappa}{2} v^4 \varphi^2 + \left( v^2 - \frac{g}{2^{\frac{n}{2}}} \varphi^n \right)^2, \\
	\label{eq:inf_potential_stb}
	V_\text{stb} (\varphi,\chi) 
	&= \frac{1}{2}c_\text{pot} V_\text{pre} (\chi) \varphi^2,
\end{align}
where we take $V_\text{pre}(\chi) = \frac{1}{2} m^2 \chi^2$ for simplicity.\footnote{
Strictly speaking, this simple choice of potential can not reproduce the Planck observational results of the scalar tilt ($n_s$) and the tensor to scalar ratio ($r$)~\cite{Ade:2015lrj}.
However, the dynamics of $\chi$ during its oscillation mainly lead to the enhancement of the perturbations with the second mechanism that we will explain in the next subsection.
Moreover, the first mechanism we will explain is independent of the detail of the first inflation.
Since the potential of $\chi$ can be approximated as the quadratic potential at the leading order during the oscillation, 
our result with the potential $\frac{1}{2} m^2 \chi^2$ can be valid even in other pre-inflation potentials,
 such as those of polynomial chaotic inflation models~\cite{Nakayama:2013jka,Nakayama:2013txa,Nakayama:2014wpa},
 which can reproduce the Planck observational results.
}
In general, we can expect the following Planck-suppressed term which is coupled with the kinetic term:
\begin{align}
	\mathcal{L}_\text{kin} = -\frac{1}{2} \left( 1- \frac{c_\text{kin}}{2} \varphi^2 \right) \partial_\mu \chi \partial^\mu \chi 
	 - \frac{1}{2}  \partial_\mu \varphi \partial^\mu \varphi + \cdots.
	\label{eq:inf_kinetic}
\end{align}
These terms given by Eqs.~(\ref{eq:inf_potential_tot})--(\ref{eq:inf_kinetic}) are naturally embeded in the supergravity (SUGRA) framework (see App.~\ref{app:sugra}).\footnote{
As described in App.~\ref{app:sugra}, in the SUGRA framework, the additional kinetic coupling term, $\frac{1}{2} \left(\frac{\kappa}{2}\varphi^2 \right) \partial_\mu \varphi \partial^\mu \varphi$, appears in the Lagrangian.
However, since $\varphi$ is small during the inflation, we can neglect the contribution from this term.
}

\paragraph*{\bf Inflation scenario and produced perturbations.}
First, we briefly explain the dynamics of our inflation model (see also~\cite{Kawasaki:1997ju, Kawasaki:2016pql,Inomata:2016rbd,Inomata:2017okj}).
During the pre-inflation, the inflaton $\varphi$, which is responsible for the new-inflation, is stabilized at the origin due to the stabilization term $V_\text{stb}$.
After the pre-inflation, the inflaton $\chi$ oscillates around its minimum and the Universe behaves as the matter-dominated Universe for a while.
The new-inflation starts when the energy related to the inflaton $\chi$, $V_\text{pre}(\chi) + \frac{1}{2} \dot \chi^2$, becomes smaller than the new-inflation energy scale, $v^4$, due to the expansion of the Universe.
After the end of the new-inflation, the inflaton $\varphi$ oscillates around its potential minimum and decays to the radiations.
In this paper, we assume that $\varphi$ decays via Planck-suppressed operators and evaluate the reheating temperature as
\begin{align}
	T_\text{R} 
	\simeq 0.1 m_\varphi^{3/2},
	\label{eq:reheating_temp}
\end{align}
where $m_\varphi$ is the mass around the minimum defined as $m_\varphi \equiv nv^2 (v^2/g)^{-1/n}$.

Next, let us move to the perturbations produced by this inflation model.
Roughly speaking, the large-scale perturbations, which are observed by CMB, are produced by the pre-inflation and the small-scale perturbations, which are the origins of PBHs, are produced by the new-inflation.
Since the inflaton $\varphi$ is stabilized at the origin by the stabilization term $V_\text{stb}$ during the pre-inflation, the large-scale perturbations are determined by the fluctuations of $\chi$ during the pre-inflation.
On the other hand, the small-scale perturbations are determined by the fluctuations of $\varphi$.
In order for PBHs to explain the DM and the LIGO events simultaneously,
the PBH mass spectrum must have peaks around $\mathcal O(10^{-13}) M_\odot$ and $\mathcal O(10) M_\odot$.
This means that the power spectrum of the curvature perturbations must have peaks at $\mathcal O(10^{12})\,$Mpc$^{-1}$ and $\mathcal O(10^6)\,$Mpc$^{-1}$.
In the double inflation model, the enhancement of the perturbations can be realized by the following two mechanisms.

The first mechanism is due to the inflection point of the new-inflation potential.
We can understand this mechanism with the slow-roll approximation.
For the perturbations produced during the new-inflation, we can approximate the power spectrum of the curvature perturbations as
\begin{align}
	\mathcal P_\zeta(k) &= 
	\frac{1}{12\pi^2} \frac{V^3_\text{new}}{{V'}^2_\text{new}} \nonumber \\ 
	&\simeq \frac{1}{12\pi^2} \frac{v^8}{\left( 2\sqrt{2}c + \kappa v^2 \varphi_k + \frac{ng}{2^{\frac{n}{2}-1}} \varphi^{n-1}_k \right)^2},
	\label{eq:pertb_slow_app}
\end{align}
where a prime denotes the derivative with respect to $\varphi$ and 
$\varphi_k$ is the value of $\varphi$ at the horizon exit of the perturbations with the scale $k$.
If $\kappa$ is negative, there is the value of $\varphi$ which makes the curvature perturbations locally maximized.
We refer to the point at which the perturbations become locally maximized as the inflection point.
Since the inflection point corresponds to the point of $V''(\varphi)=0$, the value of $\varphi$ at the inflection point can be evaluated as
\begin{align}
	&V''(\varphi_*) \simeq  \kappa v^4 + \frac{n (n-1) g v^2}{2^{\frac{n}{2}-1}} \varphi^{n-2}_* = 0 \nonumber \\
	\Rightarrow \qquad & 
	\varphi_* = \left( \frac{2^{\frac{n}{2}-1} \kappa v^2 }{n(n-1) g} \right)^{\frac{1}{n-2}}.
\end{align}
Then we can evaluate the power spectrum at the inflection point as $\mathcal P_\zeta(k_*) \simeq \frac{1}{96 \pi^2} \left(\frac{v^4}{c}\right)^2$, where $k_*$ is the scale corresponding to the inflection point.
We can see that, in order to produce the sizable amount of PBHs with this mechanism, $c \sim v^4$ is required.
Note that the perturbation peak produced by this mechanism is broad compared with that produced by the second mechanism 
because these peaks are related to the dynamics of $\varphi$ during its slow-roll.
The tilt of the power spectrum of the curvature perturbations are determined by the slow-roll parameters as
\begin{align}
	n_s -1 = -6\epsilon + 2\eta,
\end{align}
where $n_s$ is defined as $\mathcal P_\zeta (k) \propto k^{n_s-1}$
and the slow-roll parameters, $\epsilon$ and $\eta$, are defined as $\epsilon \equiv \frac{1}{2} \left(\frac{V'}{V}\right)^2$ and $\eta \equiv \frac{V''}{V}$.
The slow-roll parameters are expected to be small during the inflation.

The second mechanism is related to the Hubble-induced mass during the oscillation of $\chi$.
During the pre-inflation, the Hubble-induced mass of $\varphi$ is given by $m_\varphi^2 = 3 c_\text{pot} H^2$, where $H$ is the Hubble parameter.
Meanwhile, during the oscillation of $\chi$, the effective Hubble-induced mass of $\varphi$ is given by $m_\varphi^2 = \frac{3}{2} (c_\text{pot} + c_\text{kin}) H^2$.
If $c_\text{pot} + c_\text{kin} \simeq 0$ is satisfied, the superhorizon perturbations of $\varphi$ can avoid the damping during the $\chi$-oscillation phase because the effective mass of $\varphi$ disappears at that time.
This means that the perturbations of $\varphi$ which is superhorizon at the oscillation phase are effectively enhanced.
On the other hand, the subhorizon perturbations at the oscillation phase are not affected by the cancellation of Hubble-induced mass.
Therefore, when $c_\text{pot} + c_\text{kin} \simeq 0$ is satisfied, 
the sharp damping of the curvature perturbations appears 
at scales below the horizon scale at the oscillation phase.
Of course, the perturbations which exit the horizon well before the oscillation phase is damped by the Hubble-induced mass during the pre-inflation.
From these discussions, we can expect the sharp peak of curvature perturbations around the horizon scale at the $\chi$-oscillation phase.
Note that the peak produced by this mechanism can be sharp because the peak is related to the dynamics of $\chi$ during its oscillation (see also App.~B in~\cite{Inomata:2016rbd} for the detail explanations of this mechanism).

Suppose that the parameters of our double inflation are taken so that these two mechanisms work simultaneously.
Then the light PBHs correspond to the perturbation peak produced by the first mechanism 
and the heavy ones correspond to that produced by the second mechanism.
Thus, in our double inflation model,
the lighter PBHs have a broad spectrum while the heavier ones have a sharp spectrum. 
It is tempting to consider that 
the perturbation peak corresponding to the $\mathcal O(10^{-13})M_\odot$ PBHs is broad 
and the peak corresponding to the $\mathcal O(10)M_\odot$ PBHs is sharp.
Here, let us recall the fact that the peak of curvature perturbations which produces the $\mathcal O(10)M_\odot$ PBHs must be sharp owing to the constraints from the induced GWs and 
the CMB spectral distortion/BBN.
On the other hand, the peak which produces the $\mathcal O(10^{-13})M_\odot$ PBHs can be broad.
From above discussions, one can see that the double inflation model is appropriate for the scenario where the PBHs as DM and PBHs for LIGO events coexist.

\paragraph*{\bf Concrete parameters.}

\begin{figure}
	\centering
	\includegraphics[width=.45\textwidth]{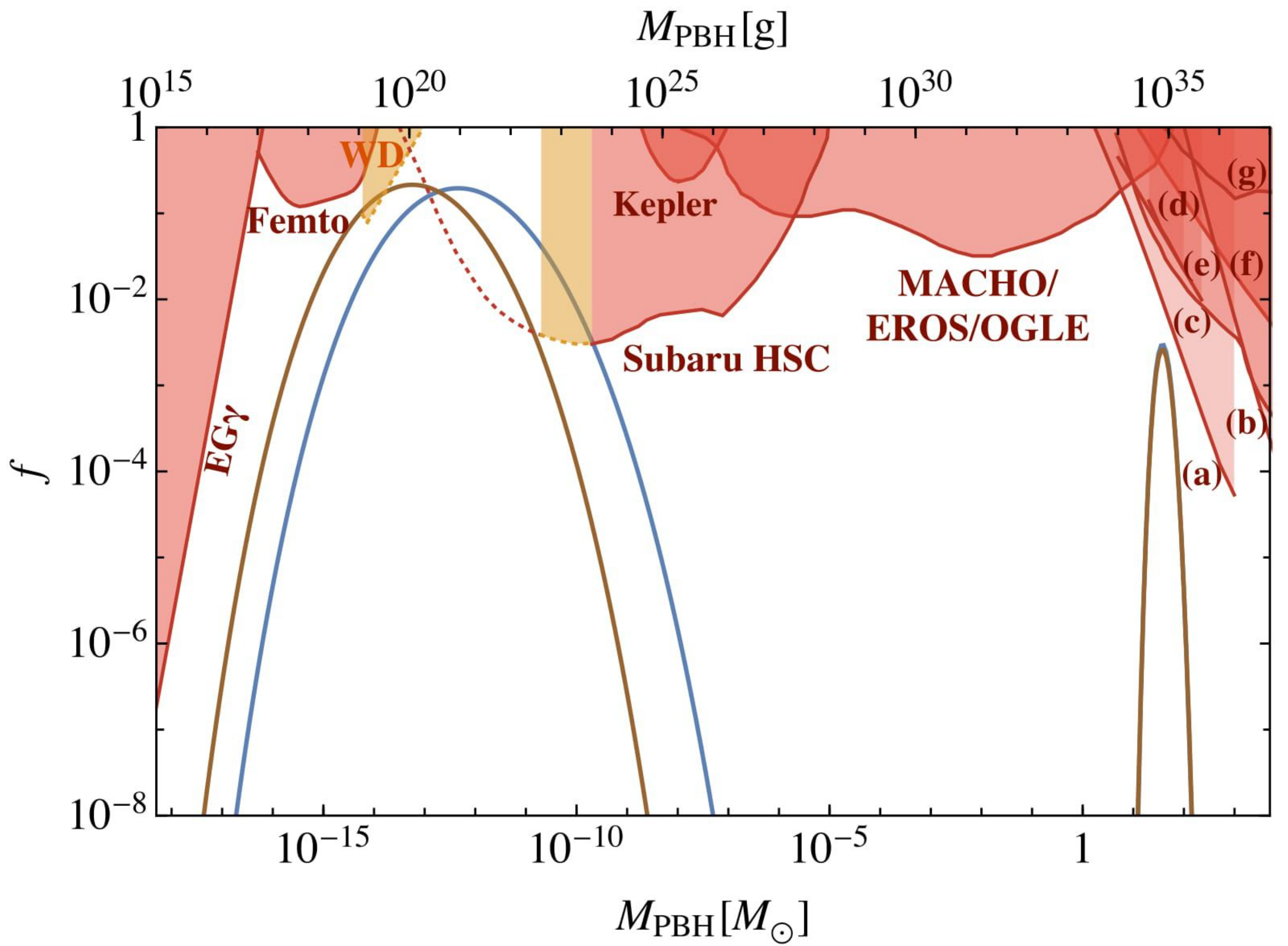}
	\caption{\small
	The PBH mass spectrums ($f(M) \equiv \frac{1}{\Omega_\text{DM}} \frac{\dd\, \Omega_\text{PBH}}{\dd \, \text{ln}\, M}$) for parameters given in Eqs.~(\ref{eq:para_case1}) (a blue solid line) and (\ref{eq:para_case2}) (a brown solid line).
	The red shaded regions show the observational constraints on the monochromatic mass function.
	The boundaries of the red shade regions correspond to $f_\text{obs}(M)$, which are defined in Eq.~(\ref{eq:f_max_def}).
	These constraints come from the observations of 
	the extra-galactic gamma-rays from the PBH evaporation (EG$\gamma$)~\cite{Carr:2009jm}, 
	the femtolensing events (Femto)~\cite{Barnacka:2012bm}, 
	the microlensing events with Subaru HSC (Subaru HSC)~\cite{Niikura:2017zjd}, 
	with Kepler satellite (Kepler)~\cite{Griest:2013esa},  
	with MACHO/EROS/OGLE (MACHO/EROS/OGLE)~\cite{Allsman:2000kg,Tisserand:2006zx,wyrzykowski2011ogle},
	the variation of CMB spectrum (a,b)~\cite{Poulin:2017bwe,Ali-Haimoud:2016mbv},
	 the radio and X-ray from accretion (c,d)~\cite{Inoue:2017csr,Gaggero:2016dpq},
	the dynamical heating of dwarf galaxies and ultra-faint dwarf galaxies (e,f)~\cite{Koushiappas:2017chw,Brandt:2016aco}, 
	and the distribution of wide binaries (g)~\cite{monroy2014end}.
	The red dotted line shows the uncertain constraint of HSC~\cite{Niikura:2017zjd} because of the wave effect~\cite{Ohanian:1974ys,Bliokh:1975wave,Bonts:1981wave}.
	The orange shaded regions show
	 the constraint from the existence of white dwarfs in our local galaxy (WD)~\cite{Graham:2015apa}, which is neglected in \emph{Case 2}, 
	and the constraint of Subaru HSC in $2\times 10^{-11} M_\odot < M_\text{PBH} < 2\times 10^{-10} M_\odot$, which is neglected in \emph{Case 1} (see text).
	}
	\label{fig:pbh_spectrum}
\end{figure}

\begin{figure}
	\centering
	\includegraphics[width=.45\textwidth]{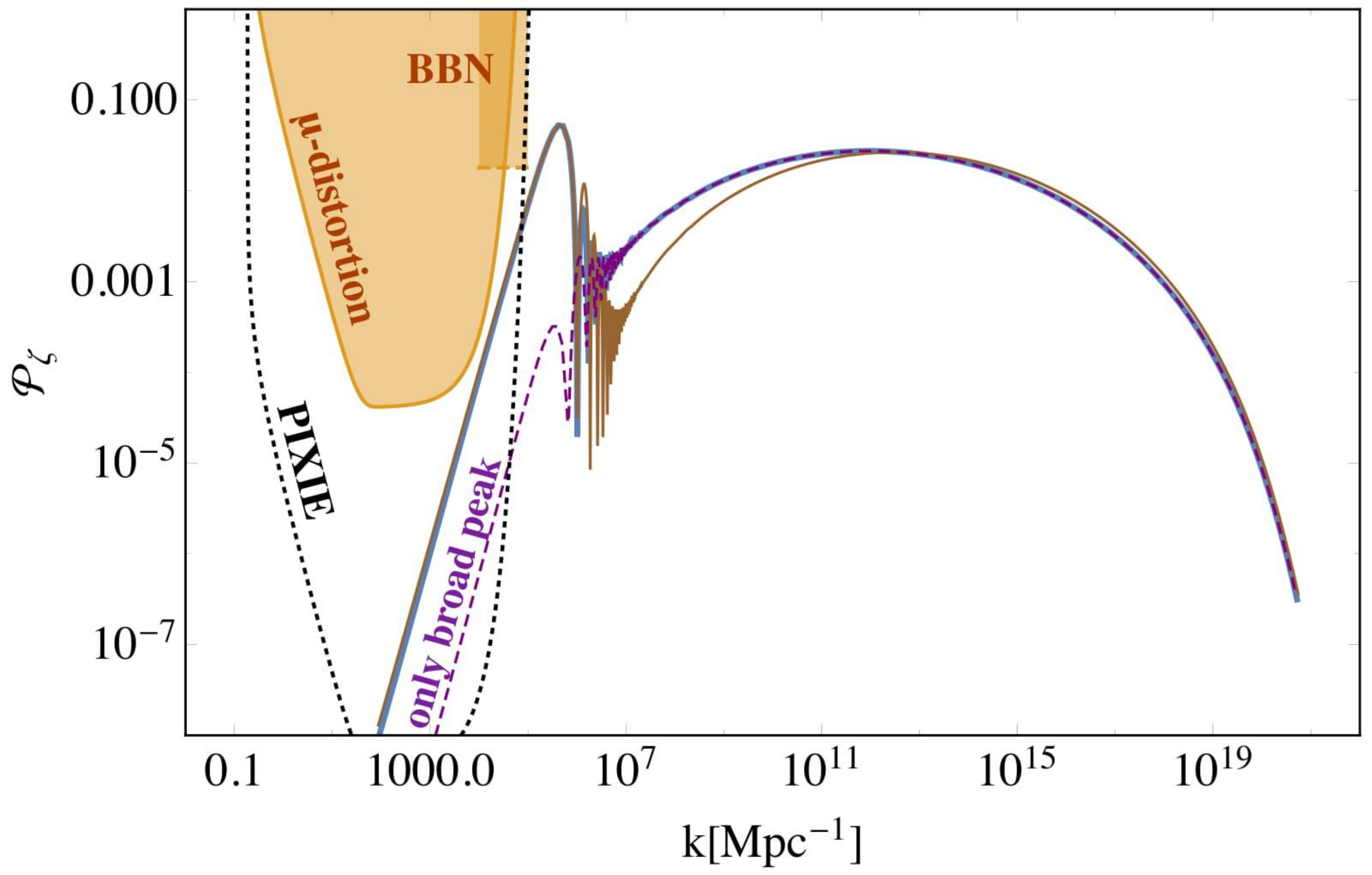}
	\caption{\small
	The power spectrums of the curvature perturbations for parameters given in Eqs.~(\ref{eq:para_case1}) (a blue solid line) and (\ref{eq:para_case2}) (a brown solid line).
	Orange shaded regions are excluded by 
	the current constraint on $\mu$-distortion, $|\mu|<9\times10^{-5}$~\cite{Fixsen:1996nj}
	and the effect on n-p ratio during big-bang nucleosynthesis~\cite{Inomata:2016uip}.	
	The black dotted line represents a future constraint by $\mu$-distortion with the PIXIE~\cite{Kogut:2011xw}, 
	$|\mu|<10^{-8}$.
	For comparison, the power spectrum of the curvature perturbations for $c_\text{kin}=0$ and the other parameters given in Eq.~(\ref{eq:para_case1}) is plotted with a purple dashed line (only broad peak).	
	}
	\label{fig:curvature_spectrum}
\end{figure}

\begin{figure}
	\centering
	\includegraphics[width=.45\textwidth]{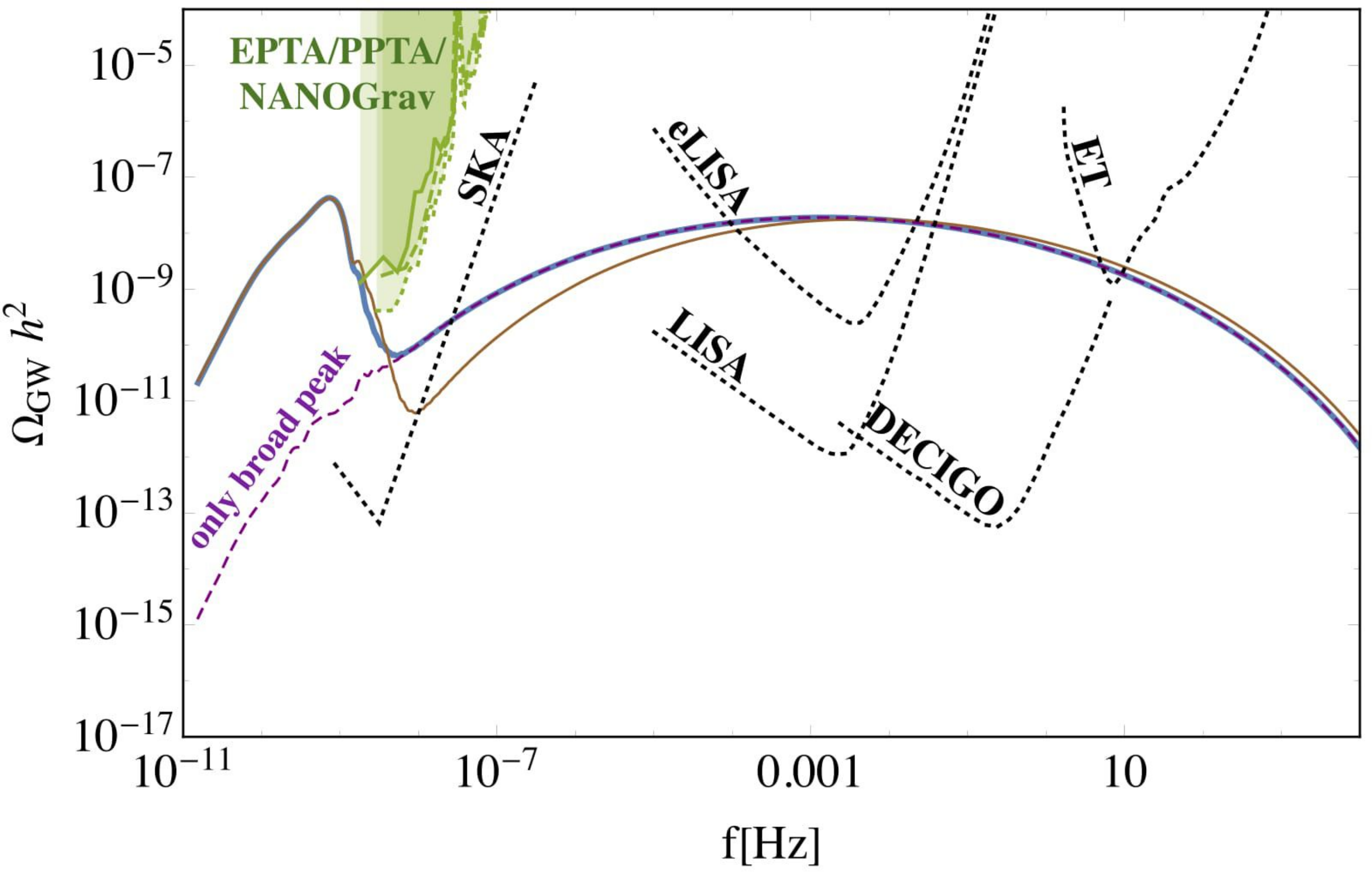}
	\caption{\small
	The induced GW spectrums for parameters given in Eqs.~(\ref{eq:para_case1}) (a blue solid line) and (\ref{eq:para_case2}) (a brown solid line).
	Green shaded regions are excluded by PTA observations with EPTA~\cite{Lentati:2015qwp}, 
	  PPTA~\cite{Shannon:2015ect}, and NANOGrav~\cite{Arzoumanian:2015liz}.
	 Black dotted lines show the prospects of the future experiments:
	 SKA (PTA)~\cite{Moore:2014lga,Janssen:2014dka},
	 eLISA~\cite{AmaroSeoane:2012km}, LISA~\cite{Sathyaprakash:2009xs,Moore:2014lga}, DECIGO~\cite{Yagi:2011wg} (space-based GW interferometer),
	 and Einstein Telescope (ET)~\cite{Sathyaprakash:2009xs,Moore:2014lga,ET_sense} (third-generation ground-based GW interferometer).
	For comparison, the induced GW spectrum for $c_\text{kin}=0$ and the other parameters given in Eq.~(\ref{eq:para_case1}) is plotted with a purple dashed line (only broad peak).
	}
	\label{fig:gw_spectrum}
\end{figure}

In this subsection, we show concrete parameter sets
with which the double inflation model produces the sizable amount of PBHs 
with $\mathcal O(10) M_\odot$ and $\mathcal O(10^{-13})M_\odot$.

Here, we show the successful parameters in the two cases, which are mentioned in Sec.~\ref{sec:constraints}.

\noindent \emph {Case 1}\,: 
We neglect the constraint of Subaru HSC below $2 \times 10^{-10}M_\odot$.
In this case, the successful parameters are as follows:
\begin{align}\label{eq:para_case1}
	n&=3, \quad v=10^{-4}, \quad \kappa=-0.39, \quad c = 1.23\times v^4, \nonumber \\ 
	g&=6.14\times10^{-3}, \quad c_\text{pot}=0.66, \quad c_\text{kin}=-0.515.
\end{align}

\noindent \emph {Case 2}\,: 
We neglect the constraint of Subaru HSC below $2\times 10^{-11}M_\odot$ and the constraint from the survival of the white dwarfs.
In this case, the successful parameters are as follows:
\begin{align}\label{eq:para_case2}
	n&=3, \quad v=10^{-4}, \quad \kappa=-1.1, \quad c = 7.54\times v^4, \nonumber \\ 
	g&=6.87\times10^{-3}, \quad c_\text{pot}=0.66, \quad c_\text{kin}=-0.86.
\end{align}

In the \emph {Case 1}, we assume that the wave effect significantly reduced the HSC constraints.
On the other hand, \emph {Case 2} is the case where the wave effect makes mild effects on the HSC constraint 
and the constraint from the white dwarfs is not valid due to the uncertainties.\footnote{
If we adopt the HSC constraint with $>2\times 10^{-11}M_\odot$ and the constraint of the white dwarfs at the same time,
it is difficult to simultaneously realize the PBHs as DM and PBHs for LIGO events in our model.
}

We plot 
 the PBH mass spectrums in Fig.~\ref{fig:pbh_spectrum},
the power spectrums of the curvature perturbations in Fig.~\ref{fig:curvature_spectrum},
  and the GW spectrums induced by the second order effect of the curvature perturbations in Fig.~\ref{fig:gw_spectrum} 
  for each parameter set (\emph{Case 1}\,: Blue solid, \emph{Case 2}\,: Brown solid).
The PBH mass spectrums satisfy $\Omega_\text{PBH,tot}/\Omega_\text{DM} = 1$.
Moreover, the mass spectrums have sharp peaks around $\mathcal O(10) M_\odot$ and their height are $\mathcal O(10^{-3})$, which reproduce the merger rate estimated by the LIGO-Virgo collaborations~\cite{Sasaki:2016jop}.
In Fig.~\ref{fig:pbh_spectrum}, we also show the constraints on the monochromatic PBH mass function by the red shaded regions.
As we discussed in Sec.~\ref{sec:pbh_as_dm}, we must be careful about the constraints on the extended mass function.
Following the analysis in Sec.~\ref{sec:pbh_as_dm}, we have checked that the mass spectrums are consistent with all the observations except 
for the ones we explicitly neglect.

From Fig.~\ref{fig:curvature_spectrum} and Fig.~\ref{fig:gw_spectrum}, 
 we see that the curvature perturbations have the peak at $k\sim 10^6$Mpc$^{-1}$, 
 which is so sharp that the perturbations are consistent with $\mu$-distortion~\cite{Fixsen:1996nj}, BBN~\cite{Inomata:2016uip}, and PTA observations~\cite{Lentati:2015qwp,Shannon:2015ect,Arzoumanian:2015liz}.
This is due to the characteristic enhancement mechanisms of the double inflation model. 
For comparison, in Fig.~\ref{fig:curvature_spectrum} and Fig.~\ref{fig:gw_spectrum}, 
we also plot the spectrums for $c_\text{kin}=0$ while keeping the other parameters in Eq.~(\ref{eq:para_case1}) unchanged (purple dashed lines).
The equation $c_\text{kin}=0$ makes $c_\text{pot}+c_\text{kin}\sim \mathcal O(1) \not\approx 0$; that is, the second enhancement mechanism does not work in $c_\text{kin}=0$.
Comparing the purple lines with the blue solid lines (\emph{Case 1}), one can confirm that the broad peaks are produced by the first mechanism and the sharp peaks are produced by the second mechanism, 
as discussed in the previous subsection.

In Fig.~\ref{fig:gw_spectrum}, we see that the induced GW spectrum is above the sensitivity of SKA~\cite{Moore:2014lga,Janssen:2014dka}, (e)LISA~\cite{Sathyaprakash:2009xs,AmaroSeoane:2012km}, DECIGO~\cite{Yagi:2011wg} and Einstein Telescope (ET)~\cite{Sathyaprakash:2009xs,Moore:2014lga,ET_sense}.
In particular, 
the frequency $f \sim$nHz, which is covered by SKA, corresponds to $\mathcal O(1) M_\odot$; and
the frequency $f\sim \mathcal O(10^{-3})$--$\mathcal O(10^{-2})$Hz, which is covered by (e)LISA and DECIGO, corresponds to $\mathcal O(10^{-13})M_\odot$ (see Eq.~(\ref{eq:pbhmass}))~\cite{Garcia-Bellido:2017aan}.
One can see that both bumps corresponding to PBHs for DM and LIGO will be probed by planned observations of GWs.
Since the induced GWs has the stochastic properties, they can be distinguished from the individual events expected at the frequency, 
such as the GWs from $\mathcal O(10^{6}) M_\odot$ BHs binaries (see~\cite{Moore:2014lga} and references therein).\footnote{
The enhancement of the sensitivity for some types of stochastic GWs is discussed in Ref.~\cite{Thrane:2013oya}.
}
Moreover, since the frequency dependence of the GW spectrum traces the scale dependence of the curvature perturbations, 
(e)LISA and DECIGO could possibly observe the sharpness of the PBH mass spectrum indirectly.
If the mass spectrum is broad enough, the stochastic GW can be observed also by ET.
Even in the case of PBHs only for all DM,
the low-frequency tail of the stochastic GW may be probed by SKA,
if the mass spectrum is broad enough.

\section{Conclusions}
\label{sec:conclusion}

In this paper, we have taken into account the uncertainties of the constraints on PBHs and discussed the scenario where PBHs are all DM.
Recently, the microlensing observation with Subaru HSC has put severe constraints on the abundance of $\mathcal O(10^{-13})M_\odot$ PBHs. 
However, the constraints have the uncertainty related to the wave effect.
The wave effect may weaken the constraints on the light PBHs with $M<2 \times 10^{-10}M_\odot$.
To clarify the impact of this uncertainty,
we have revisited PBHs as all DM in the absence of the HSC constraints on the light PBHs.
We have found that the PBHs around the sublunar mass can still make up all DM, 
and moreover, they can have the broad mass function.

Thanks to this fact, there appears the possibility that
the double inflation model can simultaneously explain PBHs as DM and PBHs as the BHs detected by LIGO.
For particular parameter sets, the double inflation model predicts the two peaks of the curvature perturbations.
The predicted peak corresponding to the light PBHs is broad and that corresponding to the heavy PBHs is sharp.
This feature of the double inflation model fits in the picture that PBHs as DM can be produced by the broad peak, while the mass function of PBHs for LIGO events has to be sharp
if there are no significant non-Gaussianity.
We have shown the concrete parameter sets with which the double inflation can simultaneously produce the light and heavy PBHs responsible for the DM and LIGO events, respectively.
We have also shown that 
the GW-frequency corresponding to the $\mathcal O(10^{-13})M_\odot$ PBHs is covered by (e)LISA and DECIGO
while that corresponding to $\mathcal O(10)M_\odot$ PBHs is close to
the sensitivity of SKA.
The energy density of GWs induced by the second order perturbations is large enough to be detected by the three experiments.\footnote{
The energy density of the induced GWs is larger than that of the stochastic GWs produced by the $\mathcal O(10) M_\odot$ PBH mergers, which is estimated in Refs.~\cite{Clesse:2016ajp,Wang:2016ana,Raidal:2017mfl}.
}
The SKA is powerful enough to probe GWs corresponding to PBHs for LIGO events 
and could probe the low-frequency tail of the broad spectrum
even if our model is not responsible for PBHs for LIGO events.
Since the induced GWs depend on the power spectrum of the curvature perturbations, 
(e)LISA and DECIGO could possibly determine the sharpness of the PBH mass spectrum.
In the future, the two experiments can test the DM scenario with $\mathcal O(10^{-13})M_\odot$ PBHs, and if the scenario turns out to be true, 
they can also help us understand the inflation model which produces the PBHs by observing the GW spectrum shape.

\section*{Acknowledgements}
{\small
\noindent
We would like to thank Yuichiro Tada for discussions at an early stage of this work
and Masahiro Takada for helpful discussions on the wave effect.
K.M.\ would like to thank G{\'e}raldine Servant for comments.
This work is supported by Grant-in-Aid for Scientific Research from the Ministry of Education,
Science, Sports, and Culture (MEXT), Japan,  No.\ 15H05889 (M.K.), No.\ 25400248 (M.K.),
No.\ 26104009 (T.T.Y.), No.\ 26287039 (T.T.Y.) and No.\ 16H02176 (T.T.Y.), 
World Premier International Research Center Initiative (WPI Initiative), MEXT, Japan 
(K.I., M.K., K.M., and T.T.Y.),
JSPS Research Fellowships for Young Scientists (K.M.),
and Advanced Leading Graduate Course for Photon Science (K.I.).
}

\appendix


\section{Double Inflation in Supergravity}
\label{app:sugra}

Here we provide an explicit realization of the Lagrangian given in Eqs.~\eqref{eq:inf_potential_tot}, \eqref{eq:inf_potential_new1}, \eqref{eq:inf_potential_stb}, and \eqref{eq:inf_kinetic} in the supergravity framework.
The model is the same as the one presented in App.~C of \cite{Inomata:2016rbd}.

\begin{table}[t]
\begin{tabular}{c||c|c|c|c|c}
&$\Psi$ & $X$ & $\Phi$ & $v^2$ & $c$ \\
\hline
$R$ charge & $0$ & $2$ & $2$ & $0$ & $2$
\end{tabular}
\caption{\small The $R$-charge assignments are shown.}
\label{tab:Rcharge}
\end{table}

In the pre-inflation sector, we have two chiral superfields;
one is used for the inflation $\Psi$, and the other is a so-called stabilizer $X$.
We require that the system respects 
a shift symmetry of $\Psi \mapsto \Psi + iA$ with $A$ being a real parameter,
which is softly broken by a holomorphic spurious parameter $m$.
As shown in Ref.~\cite{Kawasaki:2000yn}, the stabilizer prevents us from rolling down to unwanted Anti--de Sitter vacua while maintaining the $R$ symmetry not to be broken by a large field value of $\Psi$ during inflation.
In the new-inflation sector, we have one chiral superfield $\Phi$.
To achieve a sufficiently flat potential required for the new-inflation,
we assume that the underlying $R$ symmetry is $\mathbb Z_{2nR}$.
When $\Phi$ develops a VEV after the new-inflation, this discrete $R$ symmetry is broken into $\mathbb Z_{2R}$.
Furthermore, we assume that the remaining $\mathbb Z_{2R}$ is completely broken by a tiny constant term $c$ in the super potential.
This explicit breaking term is required to solve the domain wall problem 
associated with $\mathbb Z_{2nR} \to \mathbb Z_{2R}$,
since the $e$-folds of the new-inflation is less than that of the CMB~\cite{Izawa:1997df}.
Interestingly,
this term generates a linear potential for the new-inflation,
which opens up a possibility to enhance the curvature perturbation twice
as discussed in the main text.

The charge assignments of the $R$ symmetry are summarized in Tab.~\ref{tab:Rcharge}.
Let us recall here that the shift symmetry, $\Psi \mapsto \Psi + iA$,
is broken softly by a holomorphic spurious parameter $m$.
The super- and K\"ahler-potentials consistent with these requirements may be written as follows:
\begin{align}
	W =& m X \Psi - \frac{g}{n+1} \Phi^{n+1} + v^2 \Phi + c, \\
	K =& \frac{1}{2} \prn{\Psi + \Psi^\dag}^2 +\abs{X}^2 + \abs{\Phi}^2 \nonumber \\
	&+ \frac{\kappa}{4} \abs{\Phi}^4 + c_\text{pot}' \abs{X}^2 \abs{\Phi}^2 
	+ \frac{c_\text{kin}'}{2} \abs{\Phi}^2 \prn{\Psi + \Psi^\dag}^2 + \cdots,
\end{align}
where we assume $m \gg v^2$.
We have dropped terms which are not relevant for our purpose. For instance, the ellipses include a term that strongly stabilizes $X$ at the origin during the pre-inflation; \textit{e.g.,} $- c_X\abs{X}^4$ with $c_X \gtrsim 1$.
Note that, strictly speaking, 
the pre-inflation governed by these potentials does not satisfy the current observation.
However, in our paper, we are only interested in how the small-scale perturbations are generated during the new-inflation after the pre-inflation.
From this viewpoint, the above Lagrangian may be regarded as an approximate one that is valid after the onset of the $\Psi$-oscillation.\footnote{
	One may modify the potential of the pre-inflation
	at the large field value regime to accommodate the observational constraints. See ~\cite{Nakayama:2013jka,Nakayama:2013txa,Nakayama:2014wpa} for instance.
}

Now we are in a position to write down the Lagrangian relevant for the new-inflation. We rewrite the scalar components as follows: $\Psi = (\eta + i \chi )/\sqrt{2}$ and $\Re \Phi = \varphi / \sqrt{2}$.
During the chaotic inflation $\chi \gtrsim 1$, 
other scalar fields, $\eta$, $\Phi$, and $X$, are stabilized near the origin
through the Hubble induced mass terms.
After the end of the pre-inflation,
the new-inflation eventually starts when the Hubble induced mass term becomes small enough, while $\eta$ and $X$ keep stabilized by $m^2 |X|^2$ and $m^2 \eta^2$.
Hence, setting $X \simeq \eta \simeq 0$,
we may write down the relevant terms for the discussion of the new-inflation.
The potential is given by
\begin{align}
	V \simeq & v^4 - 2 \sqrt{2} c v^2 \varphi - \frac{\kappa}{2} v^4 \varphi^2
	-\frac{g}{2^{\frac{n}{2}-1}} v^2 \varphi^n + \frac{g^2}{2^n} \varphi^{2n} \nonumber\\
	& + \prn{1 + \frac{c_\text{pot}}{2} \varphi^2} \frac{m^2 \chi^2}{2},
\end{align}
where $c_\text{pot} \equiv (1 - c_\text{pot}')/2$.
The relevant kinetic terms are
\begin{align}
	\mathcal L_\text{kin} \supset
	-\frac{1}{2} \prn{1 + \frac{\kappa}{2}\varphi^2} 
	\partial_\mu \varphi \partial^\mu \varphi
	- \frac{1}{2} \prn{1 - \frac{c_\text{kin}}{2} \varphi^2}
	\partial_\mu \chi \partial^\mu \chi,
\end{align}
where $c_\text{kin} \equiv - c_\text{kin}'$.

Finally, we comment on the gravitino mass and a supersymmetry (SUSY) breaking in this model.
At the vacuum, the inflaton potential acquires a negative vacuum energy which has to be canceled out.
We assume that the cancellation occurs by the positive energy of the SUSY breaking $\mu_\text{SUSY}^4$, which may come from $W_\text{SUSY} = \mu_\text{SUSY}^2 Z$
with $Z$ being a SUSY breaking field.
The gravitino mass can be expressed as
\begin{align}
	m_{3/2} = \frac{\mu^2_\text{SUSY}}{\sqrt{3}}
	\simeq \frac{n}{n+1} v^2 \prn{\frac{v^2}{g}}^\frac{1}{n}.
\end{align}
Meanwhile, the constant term $c$ in the superpotential may also come from the same SUSY breaking effect.
In particular, models with $\vev{Z} \sim \mu_\text{SUSY}$ leads to 
$c \sim \mu_\text{SUSY}^3$.
Only for $n=3$, we have $c \sim \mu_\text{SUSY}^3 \sim v^4$,
which is required for the large curvature perturbation.
The case of $n=3$ is particularly interesting from this perspective.

\small
\bibliographystyle{apsrev4-1}

\end{document}